# All-optical control of nonlinear emission from resonant metasurfaces

Ziwei Yang[1,2], Lei Xu[3], Gabriel Sanderson[3], Akhshay Bhadwal[4], Luyao Wang[1], Katsuya Tanaka[2,5], Muyi Yang[2,5], Mingkai Liu[1], Shaun Lung[2,5], Isabelle Staude[2,5], Thomas Pertsch[2], Carl Brown[4], Mohsen Rahmani[3*], Dragomir Neshev[1*]

[1] ARC Centre of Excellence for Transformative Meta-Optical Systems (TMOS), Department of Electronics Materials Engineering, Research School of Physics, The Australian National University, ACT, 2601, Australia.

[2] Institute of Applied Physics, Abbe Center of Photonics, Friedrich Schiller University Jena, Jena, 07745, Germany.

[3] Advanced Optics and Photonics Laboratory, Department of Engineering, School of Science and Technology, Nottingham Trent University, Nottingham, NG11 8NS, United Kingdom.

[4] SOFT Group, Department of Physics and Mathematics, School of Science and Technology, Nottingham Trent University, Nottingham, NG11 8NS, United Kingdom.

[5] Institute of Solid State Physics, Friedrich Schiller University Jena, Jena, 07745, Germany.

*Corresponding author. Email: dragomir.neshev@anu.edu.au & mohsen.rahmani@ntu.ac.uk

**Nonlinear optics underpins a broad range of photonic technologies, from classical and quantum light sources to emerging nonlinear photonic neural networks. Yet, conventional nonlinear-optical devices exhibit static functionality: their transfer characteristics and emission profiles are dictated by the intrinsic nonlinear process and locked by fabrication, limiting adaptability. Here, we introduce an ultra-thin metasurface platform that enables dynamic reconfiguration of nonlinear functionality in a contact-less fashion. By leveraging all-optical control of the optical torque exerted on liquid crystal molecules infiltrating a resonant**

**metasurface, we achieve tunable polynomial nonlinear transfer functions based on third-harmonic generation process. This mechanism further allows real-time modulation of nonlinear weighting across different diffraction orders, revealing a previously unexplored interplay between mode structure and nonlinear emission. Our approach opens up a pathway toward field-programmable nonlinear photonic systems, offering unprecedented flexibility for reconfigurable nonlinear signal processing and adaptive photonic computing.**

# INTRODUCTION

Nonlinear optical processes (*1*) form a cornerstone of contemporary optics, underpinning key functionalities such as frequency conversion (*2, 3*), optical signal processing (*4*), and nonlinear wave–matter interactions (*5*) being employed in cutting-edge technologies such as lasers, optical fibers, and high precision bio-chemical imaging (*6, 7*). Efficient nonlinear interactions typically require strong light confinement or long interaction lengths, which are satisfied with relatively large nonlinear crystals in today's macroscale devices. However, the high cost of nonlinear crystals and the limited space in miniaturized devices have driven a scientific quest to generate and control nonlinear emissions with minimal reliance on material volume. Such a need has given birth to nonlinear photonics, motivating the development of resonant nanophotonic structures that enhance electromagnetic fields at subwavelength scales. In this context, metasurfaces have emerged as a compact and versatile platform for nonlinear optics, offering precise control over resonance properties, field distributions, and radiation channels.

The subwavelength resonances supported by metasurfaces lead to strong electromagnetic field confinement, which enhances light–matter interactions and enables efficient nonlinear processes such as harmonic generation (*8–15*) and frequency mixing (*16–18*). This resonant enhancement enables efficient nonlinear emission from ultra-thin devices and allows tailoring of spectral, spatial, and polarization responses. Furthermore, the same resonant nature that enables strong nonlinear enhancement also renders metasurfaces highly sensitive to external perturbations, providing a natural pathway toward dynamically reconfigurable nonlinear functionality (*19–21*).

Actively tunable metasurfaces (*22, 23*) have therefore emerged as a central platform for next-generation flat optics, offering unprecedented dynamic control over light–matter interactions at subwavelength scales. Unlike static devices whose functionality is fixed after fabrication, active metasurfaces enable real-time modulation of optical responses and have been widely explored for dynamic beam steering (*24–27*), focusing (*28, 29*), programmable holography (*30, 31*), and optical information processing (*32–34*). However, the effectiveness of a tunable metasurface ultimately depends on the availability of robust mechanisms that enable reversible, high-speed, and low-loss modulation of its optical response.

The tunability of metasurfaces originates from external stimuli that modify the material refractive index (*35–37*), the ambient environment (*38, 39*), or the structural geometry (*40–42*). Such perturbations alter the resonance conditions of the metasurface and consequently modulate its optical responses. Owing to the strong electromagnetic field confinement within metasurfaces, even small variations in these parameters can lead to substantial changes in resonance frequency, linewidth, and field distribution. This sensitivity enables diverse tuning mechanisms based on electrical (*43–47*), thermal (*48–50*), or optical excitation (*35, 51–53*). When applied to nonlinear metasurfaces, these tuning mechanisms provide a means to dynamically control nonlinear emission intensity, spectral position, and efficiency through the modulation of resonant field enhancement.

Despite these advances, the dynamic modulation of nonlinear metasurfaces remains a formidable challenge. Conventional tuning approaches, including thermal, electrical, and mechanical actuation, suffer from intrinsic limitations—thermal methods are slow, electrical schemes require electrodes that may introduce losses or fabrication complexity, and mechanical modulation lacks scalability. More fundamentally, although spectral resonance shifts can modulate nonlinear conversion efficiency, such tuning does not fundamentally reconfigure the spatial field profile or mode overlap that governs nonlinear source generation. As a result, conventional approaches primarily enable amplitude modulation rather than true functional reconfiguration of nonlinear pathways. Achieving dynamic control over nonlinear functionality therefore requires in situ reshaping of the local electromagnetic field distribution and multipolar interactions. Identifying a contactless and reversible tuning mechanism capable of directly modifying the modal environment, while simultaneously influencing both linear resonances and nonlinear field enhancement, is therefore of critical fundamental importance.

Liquid crystals (LCs) provide a uniquely versatile solution to this challenge. Their long-range molecular order provides large and continuously adjustable birefringence, enabling smooth modulation of the refractive index across visible and near-infrared wavelengths. Owing to their fluidic nature, LCs can conformally surround nanostructures and efficiently interact with the confined near fields of metasurfaces. As a result, small orientation-induced changes in the liquid-crystal director can produce pronounced modifications of the resonant response. This intimate coupling makes liquid crystals highly compatible with resonant metasurfaces, enabling effective tuning of both linear (*54–56*) and nonlinear (*20*) optical responses. However, the tunability of metasurface–LC systems generally depends on external control mechanisms, such as wire-connected electrodes for electrical biasing, localized thermal heating, or chemical engineering of prealignment layers to enforce director reorientation. These approaches introduce structural complexity and practical constraints that limit the full reconfigurable potential of liquid crystals in hybrid photonic platforms.

In this work, we employ an optical-torque-driven tuning mechanism that realizes purely optical reconfiguration of metasurface–LC systems, in a contactless fashion. By encapsulating a dielectric metasurface within a Teflon-aligned LC cell, we exploit the reorientation of LC molecules induced by the polarization-dependent optical torque of the excitation beam. This mechanism provides dynamic reversible tuning of the local refractive-index up to $\Delta n = 0.2$ anisotropy without requiring heating or electrodes. Using this purely optical stimulus not only enables us to modulate the linear optical properties of the system, but also provides the opportunity for a unique nonlinear polynomial modulation, i.e. third-harmonic generation (THG) in our case, arising from optical-torque-induced tuning of the resonant mode. It modifies the local field enhancement and, consequently, the effective nonlinearity of the system. Furthermore, we observe power redistribution among diffraction orders in the nonlinear regime, establishing a link between mode symmetry, nonlinear emission, and far-field beam shaping.

This study establishes a new theoretical and experimental paradigm for all-optical control of nonlinear metasurfaces. Fundamentally, it reveals a higher-order relationship between resonant modulation and effective nonlinear susceptibility, demonstrating that external optical excitation can directly govern the strength and symmetry of nonlinear emission. Moreover, the developed framework introduces an external-force-driven nonlinear temporal coupled-mode theory, which quantitatively describes the dynamic interplay between optical torque, resonant mode evolution, and harmonic generation. By controlling nonlinear dynamics via the optical torque, this work bridges the gap between microscopic reorientation processes and macroscopic field modulation. Such a framework not only advances the theoretical understanding of tunable nonlinear photonics but also enables practical implementations of contactless and reversible spatial light modulation, optically driven multifunctional meta-devices, neuromorphic photonic architectures, and adaptive nonlinear imaging.

# RESULTS

## Conceptual demonstration of tunable mechanism

The conceptual illustration underlying contactless, reversible, and broadband light manipulation via optical torque is presented in Fig. 1. As can be seen in Fig. 1(A), our system employs LC as an anisotropic medium for an all-optical tuning. By adjusting the orientation of the nematic LC director, one can effectively change its refractive index tensor. Distinguished from traditional LC tuning methods, we employ optical torque induced by linearly polarized light in the nematic LC system which is referred to as polarization-induced optical torque (PIOT) (*57*). This technique was introduced in 1980's (*58, 59*) but due to relatively lower performance as compared with electrical or heating tools, it has not been employed in pure liquid crystal systems for decades. In this work, we exploit the capacity of meta-optics to enhance nanoscale light-matter interactions via multipolar electric and magnetic resonances, and merge this concept with dielectric nanoscale resonators to achieve a versatile light modulator in both the linear and nonlinear regime.

To realize this mechanism, all-dielectric metasurfaces composed of silicon nano-resonators that support strong Mie-type electric and magnetic resonances, are employed and excited by a broadband white light source. A femtosecond pump pulse, spectrally distant from the resonant wavelength, at near-infrared was used to drive the LC. To mitigate heat accumulation, the metasurface is fabricated entirely from all-dielectric materials. The pump beam applies an external optical force to the LC medium, causing a reorientation of the LC director. This rotational effect, arising from the optical force, is referred to as optical torque.

To implement PIOT in the LC-metasurface system, we first prepare a homeotropic alignment layer to pre-align the LC director perpendicular to the metasurface plane (*60*). Upon the application of an external force, specifically an electric field by the pump, the LC director undergoes reorientation toward a direction parallel to the metasurface plane. This reorientation modifies the anisotropic environment surrounding the metasurface, thereby altering its optical response. By characterizing the far-field optical transmission spectra, we observe a clear resonance shift from the unexcited resonance state of the metasurface driven by the LC reorientation due to PIOT, as illustrated in Fig. 1(B).

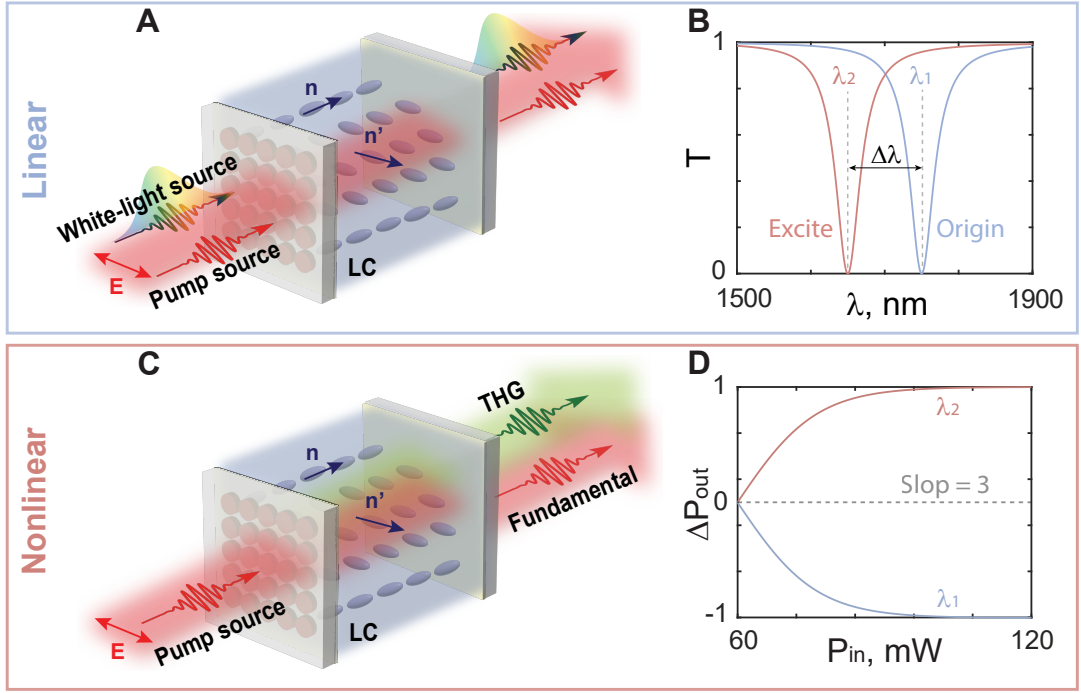

**Figure 1**: **Conceptualized illustration of linear and nonlinear optical response in liquid crystal (LC) tunable metasurfaces** (**A**) Schematic of linear response influenced by LC rotation driven by polarization-induced optical torque. (**B**) Transmittance spectrum of resonance shift due to LC rotation. (**C**) Demonstration of the third harmonic generation (THG) from a silicon metasurface controlled by liquid crystal rotation. (**D**) Positive and negative deviations of the third harmonic output caused by the resonance shift.

Most importantly, in addition to the linear-regime optical modulation, our light-driven dynamic process provides a unique modulation opportunity in the nonlinear regime. It enables intriguing large self-modulated behavior in the nonlinear regime that has not been reported with other nonlinear metasurface system to date. As illustrated in Fig. 1(C), here, a infrared pump source has been used both as the LC-driving field, and as the excitation source for the nonlinear optical process. When this pulse excites the centrosymmetric Si-based metasurface, THG is observed from the metasurface. Importantly, the LC orientation is not static and strongly dependent on the driving pump field intensity. As a result, the metasurface resonances at both the pump and harmonic wavelengths experience a dynamic, intensity-dependent spectral shift, enabling an actively tunable platform that further enriches the overall nonlinear responses based on the resonant excitation and interference effects.

When our metasurface supports a resonance at $\lambda_1$, distinct from the linear process, when the fixed pump wavelength is initially set at the resonance position $\lambda_1$, the resonance gradually shifts away from $\lambda_1$ as the pump energy increases. On the contrary, if the pump wavelength is $\lambda_2$, the resonance progressively shifts toward $\lambda_2$ with increasing pump energy. This dynamic spectral tuning introduces a polynomial nonlinear transfer function, depending on the evolving resonance position. Under conventional conditions, the TH energy follows a cubic relation with input pump power, yielding a straight line with a slope of 3 on a logarithmic plot. However, when incorporating the dynamic detuning effect on a logarithmic scale, the nonlinear response exhibits a pair of opposing deviations in slope at $\lambda_1$ and $\lambda_2$ (with a straight slope indicating no deviation), as shown in Fig. 1(D), where $\Delta P_{out}$ denotes the deviation of the output power from that obtained without dynamic detuning. This effect reflects a gradual, higher-order nonlinear modulation superimposed on the conventional THG process.

**Polarization-induced optical torque in LC-infiltrated metasurfaces**

To gain an in-depth insight into the physics process in a realistic scenario, we first design a crystalline silicon metasurface on a quartz substrate. The unit cell schematic is shown in Fig. 2(A). In the simulation, the period of the unit cell is $P = 955$ nm, the height of the nano-cylinder $h_{Si} = 245$ nm, and the radius is $R = 350$ nm. Treating the quartz substrate as semi-infinite, we simulate the static transmittance spectrum under a normally incident plane wave, as depicted by the gray dashed line in Fig. 2(B) and labeled "Air". This transmittance spectrum reveals two fundamental resonances at 1536 nm and 1676 nm. Considering the realistic alignment layer of the LC cell, a Teflon layer $h_t = 150$ nm is subsequently introduced as a prealignment layer. Additionally, an LC (E7) layer is added on top, as the surrounding medium, with a thickness of $h_{LC} = 2.85$ μm. Due to the prealignment governed by the Teflon chemical anchoring force, the LC molecules' director will present a homeotropic state, with the LC director perpendicular to the surface. The top cover layer is also simulated as semi-infinite and coated with an identical Teflon layer, with the same thickness as the bottom one. After incorporating all parameters into the model, the simulated transmittance is represented by the blue curve. We observe a clear resonance overlap between two resonances due to the surrounding refractive-index change.

To demonstrate the dynamic behavior of the PIOT-driven LC metasurface in full three dimensions under varying pump powers, we use a continuous LC layer orientation direction modeling approach. Specifically, we first assume that the LC near the metasurface will not be rotated by the optical torque, as the surface anchoring force is stronger than the optical force. Only the LC above the metasurface can be reoriented by the optical force. To provide a qualitative description and solution, we simplify the governing equation as follows (*57*), referring to it as the Optical Freedericksz Transition. (Detailed equation derivations can be found in text S1.)

$$K\frac{d^2\theta}{dz^2} + \frac{\varepsilon_0\Delta\varepsilon}{4}\sin 2\theta |E_x|^2 = 0. \tag{1}$$

Here, the $K$ denotes the elastic constant of LC, and $\Delta\varepsilon = \varepsilon_\parallel - \varepsilon_\perp = 0.72$ for E7, where the $\varepsilon_\parallel$ and $\varepsilon_\perp$ are the optical dielectric constants parallel and perpendicular to the LC director, respectively. To reorient the LC molecules via the PIOT, the applied optical torque must ≫ the elastic torque. In Eq. 1, the $\frac{\varepsilon_0\Delta\varepsilon}{4}|E_x|^2$ term represents the effective optical torque acting on the LC under *x*-polarized light. We also define the bottom surface of the silicon cylinder as the origin in Cartesian coordinates, applying the boundary conditions $\theta(h_{Si}) = \theta(h_{LC}) = 0$. Then, we solve the LC rotation as a function of distance, as shown in Fig. 2(C). In the figure, the continuous LC rotation angles are calculated according to different pump power levels with coded colors, illustrating a smooth, continuous rotation along the light propagation direction. Here, the *x*-axis represents the height of the LC cell, while the *y*-axis indicates the LC rotation angle in the $x - z$ plane, in radians. To benchmark the required power level, we further provide an order-of-magnitude estimate of the incident intensity for optically driven reorientation based on a small-angle torque balance, $I_{th} \approx ncK/\Delta\varepsilon h_{LC}^2 \approx 0.896$ kW/$cm^2$, which is consistent with the experimental pumping

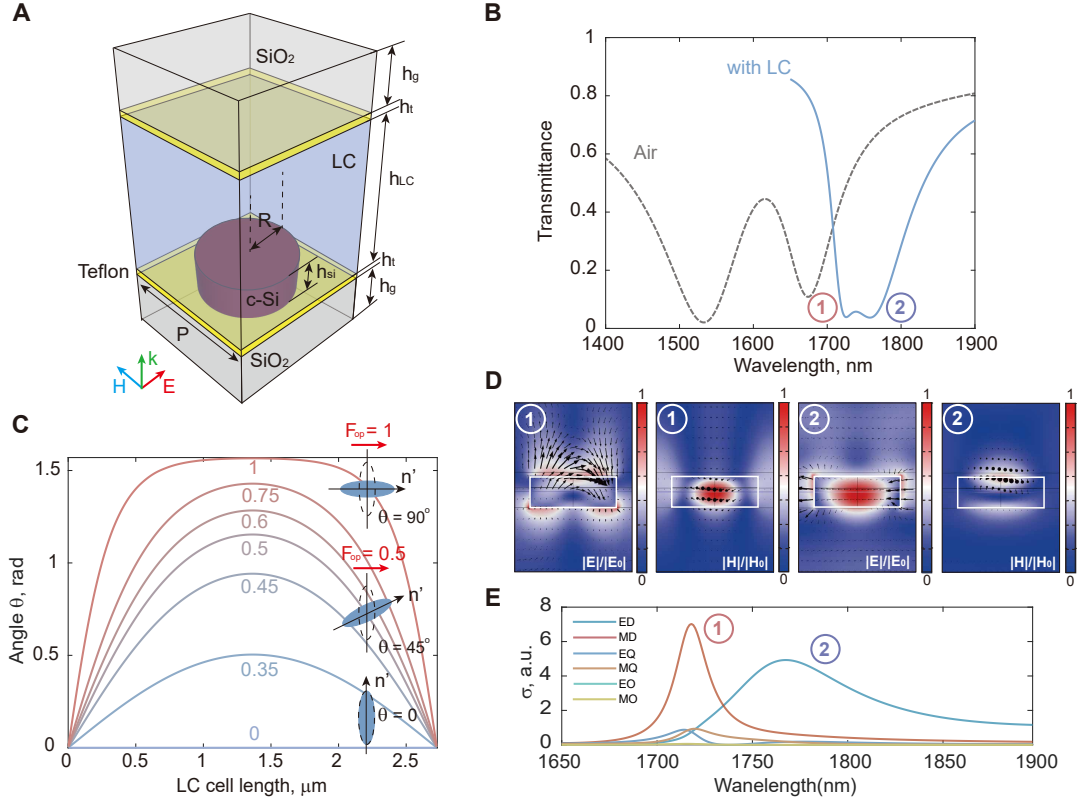

**Figure 2**: **Theoretical optical response prediction of silicon-based metasurface infiltrated with E7 LC.** (**A**) 3D view of the metasurface unit cell, the prealignment material Teflon is coated on both sides of the LC cell to provide a hard-boundary condition for homeotropic alignment. By giving the light excitation from the back of the metasurface, we designed parameters of the unit cell as $P = 955$ nm, $h_t = 150$ nm, $h_{Si} = 245$ nm, $h_{LC} = 2.85$ $\mu$m, $R = 350$ nm, and $h_g$ is infinity. (**B**) The simulated transmittance of the designed metasurface before and after LC infiltration. Two modes are labeled by number 1 (light red) and 2 (light purple). Calculated LC space-inhomogeneous rotation angles with different normalized electric fields are shown in (**C**). Maximum LC rotation angle in rad is 1.57($\approx 90°$). (**D**) Normalized magnitudes of electric and magnetic fields for two modes with LC infiltration. Field analysis using multipolar decomposition in (**E**) shows that mode 1 (resonance 1) is the magnetic dipole dominant and mode 2 (resonance 2) is the electric dipole dominant.

conditions (details in text S1).

It is worth noting that the total LC thickness is chosen to be 3 $\mu$m, as our simulations confirm that the metasurface near field scattering response is negligibly affected by refractive index variations occurring above 1 $\mu$m. Using the calculated LC director angle distribution, we derive the effective anisotropic refractive index by applying a rotation matrix through Euler angle transformation. The general description of crystalline axes rotation is shown as Eq. 2,

$$\hat{n} = \hat{\rho}(\alpha, \theta) \begin{bmatrix} n_o & 0 & 0 \\ 0 & n_o & 0 \\ 0 & 0 & n_e \end{bmatrix} \hat{\rho}(\alpha, \theta)^{-1} \qquad (2)$$

In Eq. 2, $\hat{\rho}$ is the rotation matrix, $\alpha$ represents the rotated angle in the x-y plane, and $\theta$ denotes the angle between the LC director and the $z$ axis (details in text S1).

Building on the anisotropic refractive index distribution derived from the LC rotation profile, we further investigate how these spatial anisotropies impact the mode response. As we discussed in previous work (*38*), the optical responses of these two resonances differ due to the distinct spatial distributions of their corresponding modes. As plotted in the normalized electric and magnetic fields in Fig. 2(D), these modes reveal that they are no longer perfectly localized inside the cylinder compared with air, owing to the infiltration of E7 LC in the surroundings. To rigorously classify the resonances, we perform a multipolar decomposition analysis. The results, presented in Fig. 2(E), indicate that the resonance labeled as 1 (red curve) is a magnetic-dipole-dominated mode, whereas resonance 2 (blue curve) exhibits a dominant electric-dipole character.

## Linear experimental characterization

Guided by our simulations, we fabricated a Si metasurface using standard electron-beam lithography and etching processes. Following fabrication, a Teflon layer was coated onto the structure, as shown in the scanning electron microscopy (SEM) image in Fig. 3(A). Details of the coating process are provided in Materials and Methods (S2). In this SEM image, the rough background corresponds to the Teflon layer, which is uniformly coated on the quartz substrate and exhibits nanometer-scale particulates. This coating plays a crucial role in aligning the LC molecules in a homeotropic orientation state.

Before assembling the LC cell, we characterized the transmittance of the metasurface, as shown by the gray curve labeled "Air" in Fig. 3(B), which agrees well with the simulated response of the static design. We then assembled the LC cell and infiltrated it with E7, using standard procedures, as described in (*38, 55*) (also see Materials and Methods, S2). Once the LC molecules were fully reoriented by the pre-alignment layer, we assessed the uniformity of the LC alignment using a cross-polarized optical microscope (details in S2). Based on these observations, we can confirm that the LC molecules are homeotropically aligned within the cell.

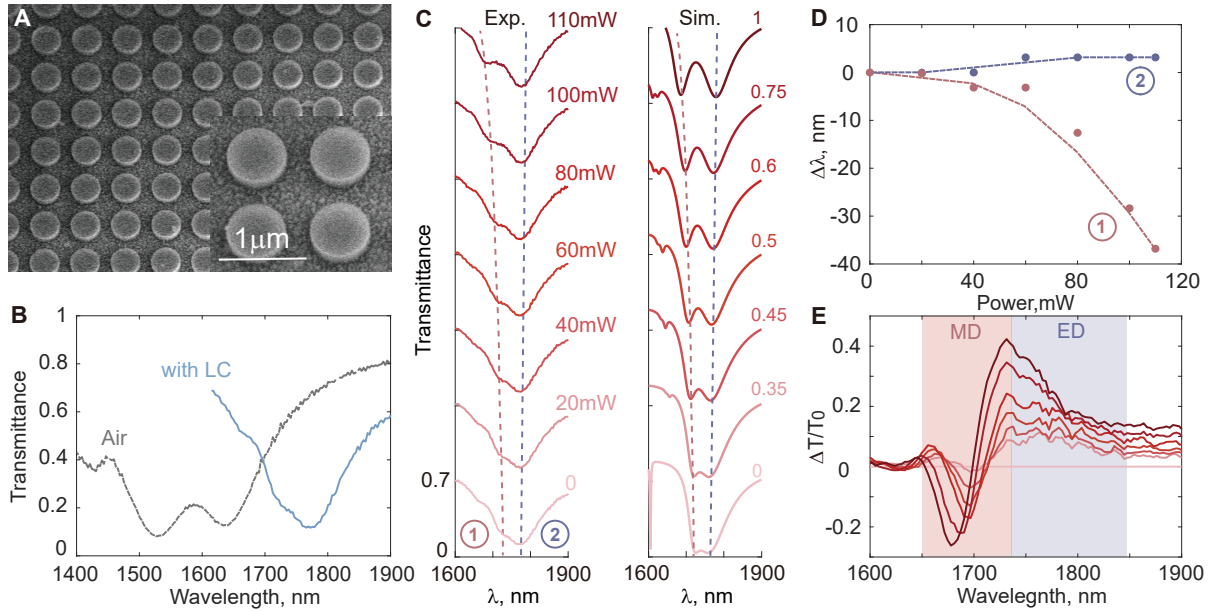

Figure 3: **Fabrication image and linear optical response experimental characterization.** (**A**) SEM images showing a bare metasurface coated with a thin Teflon layer, where the Teflon molecules are uniformly distributed around the nanodisk array. (**B**) The measured transmittance of the metasurface before and after LC infiltration. Comparing the transmittance results from both experiment and simulation in (**C**), the resonances 1 and 2 illustrate good consistency as the laser average power increases. By extracting the experimental data along the red and blue dashed lines in (**C**), we track the shifts of the two resonances, as shown in (**D**). In addition, the power-dependent modulation of transmittance under resonant wavelength is presented in (**E**), which shows the ratio of transmittance difference at the initial (1716 nm) and final (1685 nm) resonance position.

After infiltration, the transmittance spectrum, shown as the light blue curve in Fig. 3(B),

was measured under normally incident, linearly polarized light (polarization aligned along the metasurface x-axis), with pump and probe beams co-polarized. The measured response matches well with the simulation predictions (see experimental setup in S3). In this configuration, the magnetic and electric dipole resonances partially overlap, consistent with the simulated results, although the measured spectrum exhibits increased intrinsic losses. Following this static characterization, we proceed to investigate the dynamic modulation of the optical response induced by PIOT.

To enable the modulation by PIOT, we first employ a narrowband pulse laser centered at 1450 nm, with a pulse duration of 150 fs and a repetition rate of 8 MHz, as the pump beam. The broadband white light source used in the static measurements now serves as the probe beam. This narrowband pump beam is spatially expanded using an optical expander to ensure uniform rotation of the LC molecules during the pumping period. The pump intensity is sufficiently high and the beam size fully covers the transmittance measurement area, ensuring that most LC molecules within this region are effectively illuminated. Starting from the static state (labeled as 0) and gradually increasing the average pump power up to 110 mW, we record the transmittance spectra step by step, as shown in Fig.3(C). As the power increases, the magnetic dipole-dominated resonance (mode 1) exhibits a pronounced blue shift, while the electric dipole resonance (mode 2) undergoes only a slight red shift. This trend is consistent with the simulated transmittance spectra at various normalized stages of LC rotation, shown on the right-hand side of Fig.3(C), where the numbered labels correspond to the LC orientations illustrated in Fig. 2(C).

We attribute the observed resonance shifts to the distinct field distributions of the two modes. Mode 1 is primarily sensitive to the *z*-component of the refractive index, which changes from 1.7 to 1.5 during LC reorientation. Its electric field is localized above and below the structures, allowing it to strongly interact with the modulated refractive index environment. In contrast, mode 2 is sensitive to the *x*-index component, which changes from 1.5 to 1.7. However, its field is predominantly confined to the inside and both sides of the pillars. These regions are stable or filled with the static Teflon layer, thus limiting the overall index change. By tracking the resonance positions in Fig. 3(C), we extract the experimental shifts by identifying the local minima of each mode. The red and blue dots represent the experimental data points for mode 1 and mode 2, respectively. Their corresponding trends are highlighted with dashed lines of matching colors. Mode 1 exhibits a pronounced spectral shift of approximately $\Delta\lambda \approx 36$ nm, whereas mode 2 shows only a minor shift of about 3 nm. To further confirm that the observed resonance shifts are driven by PIOT rather than thermal accumulation, we performed control measurements using a probe beam with perpendicular polarization. If the shifts were caused by heating, a uniform change in the LC refractive index would result in resonance shifts regardless of probe polarization. However, as shown in S4, no spectral shift was observed in either resonance under increasing pump power in this situation, supporting the conclusion that the modulation arises from PIOT.

Next, we plot the corresponding transmittance modulation ratio with different power, $\Delta T/T_0$, shown in Fig. 3(E). The results reveal that modulation becomes evident for an incident power above 60 mW. Using this value as the pump power, we estimate the experimental threshold for the rotation of LC in our cell to be around 0.937 k$W/cm^2$ in terms of average intensity. This result indicates that observable PIOT-induced optical modulation occurs only when the PIOT exceeds the elastic force of the LC medium. As shown in the figure, the transmittance at 1685 nm decreases relative to the initial state, while an increase is observed at 1716 nm, reflecting the separation of the two resonances under optical torque. Therefore, these results demonstrate clear resonance modulation driven by PIOT within the linear regime. Once the applied optical torque exceeds the intrinsic elastic force, the LC molecules rotate to the targeted orientation, enabling dynamic resonance tuning.

This PIOT-based approach enables non-contact dynamic modulation of the nanostructures' optical properties and offers the potential for spatially non-uniform control of LCs. By tailoring complex incident fields, it holds promise for realizing high-dimensional, dynamic light–matter interactions. Such capability opens a new pathway for designing reconfigurable photonic systems with tunable functionalities that are inaccessible through conventional electro-optic or thermal methods.

## THG nonlinear experimental characterization

In the nonlinear optics regime, the driving laser pulses not only perturb the system but can also serve as the source for generating the third harmonic signals. Building on this, we further investigate the dynamic nonlinear light-matter interaction in a rotating LC nanostructure under high-power laser excitation. Using the same pulse laser as in the linear measurements, we set the pump wavelength at the fundamental resonance regime and replaced the long-pass filter with a short-pass filter to prevent direct high-power damage from the pump. The probe lamp was removed from the system, and all THG signals were recorded using a spectrometer (see Supplementary S3 for details).

In our design, the femtosecond laser first interacts with the backside of the measurface before propagating through the LC domain. This configuration helps avoid beam distortion caused by LC inhomogeneity, which could otherwise destabilize the incident light and interfere with the nonlinear response. By increasing the power of the incident light, we also manipulate the rotation of the LC, thereby shifting the resonance, as previously described. As a result, the system exhibits a dynamic nonlinear efficiency tuning compared to a static metasurface with a fixed resonance. We also verified the origin of the nonlinear signal by performing control measurements in regions without the Si metasurface. In these areas, THG was only observed when the pump power exceeded approximately 300 mW, indicating that the metasurface plays a dominant role in the THG.

We set the driving source at wavelengths of 1685 nm, 1716 nm, and 1756 nm, respectively. These three wavelengths correspond to three distinct resonance conditions that illustrate a dynamic nonlinear enhancement. At 1685 nm, the magnetic dipole resonance gradually shifts into the pump wavelength; at 1716 nm, the resonance moves out of the pump; and at 1756 nm, the nonlinear generation is attributed to the static electric dipole mode. The measured intensities from the spectrometer are presented in Fig. 4(A-C), with blue, red, and yellow data corresponding to fundamental wavelengths (FW) of 1685 nm, 1716 nm, and 1756 nm, respectively. For each wavelength, the experiment was repeated three times, and the results are shown as data points with error bars. Notably, the TH data at the fundamental wavelength, FW = 1685 nm, follow a curvature trend above a slope of 3 on a logarithmic scale. This trend indicates a dynamic enhancement of THG. In contrast, the data for the FW at 1716 nm show a curve below this slope, suggesting a dynamic decrease of the nonlinear response. These bending curvatures deviate from the expected cubic relationship between input and output power due to the dynamic involvement of the resonance. As the resonance shifts into the pump wavelength, the enhancement intensifies, leading to the observed upward bending curvature phenomenon, shown in Fig. 4(A). Conversely, as the resonance gradually moves out of the pump wavelength, the resonance-induced enhancement diminishes, resulting in a downward-bending curvature, shown in Fig. 4(B). As a reference, the case FW = 1756 nm corresponds to a static resonance condition. Figure Fig. 4(C) that the measured TH intensity scales cubically with pump power, yielding a slope of approximately 3 on the logarithmic scale, consistent with third-order nonlinear generation.

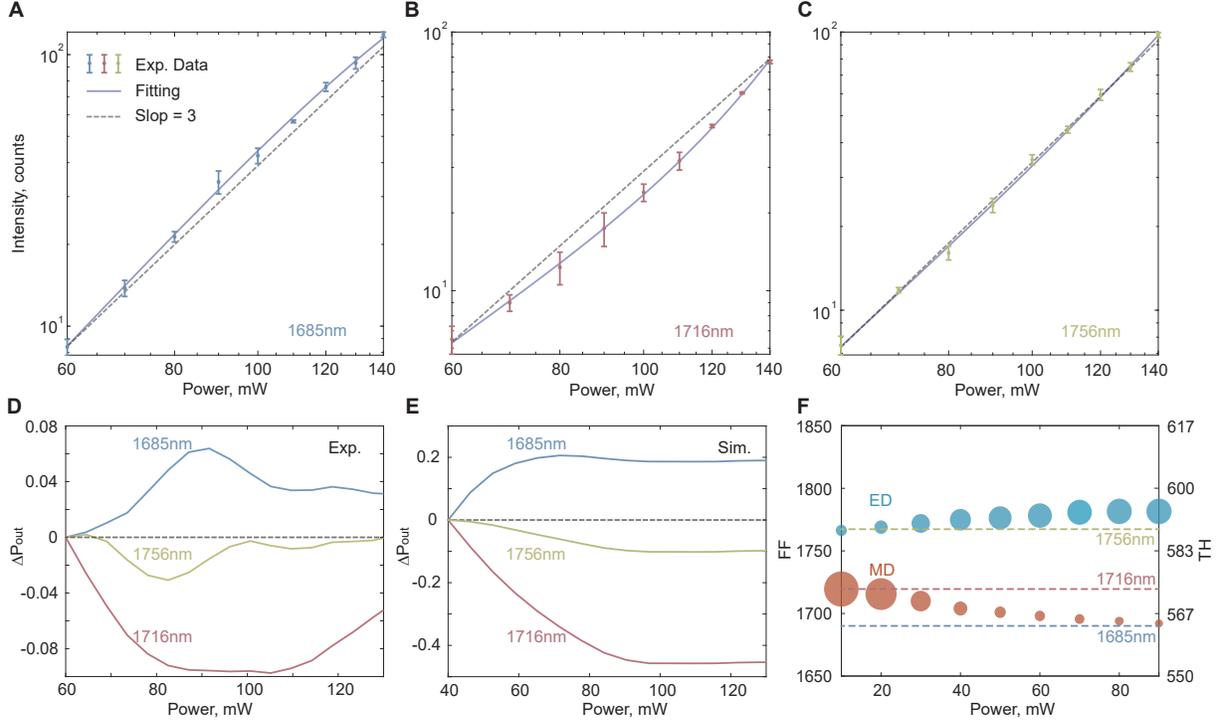

**Figure 4**: **Third harmonic nonlinear optical response characterization.** Log-log−scaling power dependent THG plot for the fundamental wavelength (FW) at 1685 nm (**A**), 1716 nm (**B**), and 1756 nm (**C**) separately. Experimental (**D**) and simulated (**E**) cubic bending deviations with three picked (**A, B, C**) wavelengths. (**F**) shows the influence of fundamental resonance shift with respect to THG efficiency according to multipolar decomposition analysis. Here FF and TH denote the fundamental and third-harmonic frequencies, respectively.

These changes in THG power dependence can be interpreted as PIOT modulation of the resonances supported by the metasurfaces, and as a result, the nonlinear response of the metasurface. Specifically, the LC reorientation, which depends on the average laser power, modifies the effective $\chi^{(3)}$ of the metasurface. To accurately capture the intensity correlation, we extended the phenomenologically established nonlinear power law to include the dependence of the nonlinear susceptibility on the average power, $\chi^{(3)} = \chi^{(3)}(P_\omega^{\text{ave}}) \approx \chi_0^{(3)} + \chi_1^{(3)} P_\omega + \chi_2^{(3)} P_\omega^2 + \cdots$. Here, $\chi_1^{(3)}$ and $\chi_2^{(3)}$ represent the effective third-order nonlinear susceptibility dependent on the incident average power, derived from a Taylor expansion of the original $\chi^{(3)}$ (Detailed expansion and susceptibility definitions are in S5).The corresponding expression for the TH power is then given below:

$$P_{3\omega} = \chi_0^{(3)2} P_\omega^3 + 2\chi_0^{(3)} \chi_1^{(3)} P_\omega^4 + \left(\chi_1^{(3)2} + 2\chi_0^{(3)} \chi_2^{(3)}\right) P_\omega^5 + 2\chi_1^{(3)} \chi_2^{(3)} P_\omega^6 + \chi_2^{(3)2} P_\omega^7 \qquad (3)$$

By assigning wavelength-dependent susceptibility values, we generate the purple fitting curves shown in Fig. 4(A-C).

To quantify the power deviation of each curvature from the slope of 3, we extract the output power difference, $\Delta P_{out} = log_{10} P_{3\omega}^{\text{PIOT}} - log_{10} P_{3\omega}^{\text{static}}$, from both the experimental data, Fig. 4(D), and the simulation results, Fig. 4(E). The experimental results exhibit the same general trend as the simulation; however, the simulated response shows a similar scale deviation at lower power, occurring at approximately 40 mW, compared to 60 mW observed in the experiment. For FW = 1685 nm, a positive deviation is observed, attributed to the THG enhancement as the resonance shifts into the pump wavelength. In contrast, as we explained above, FW = 1716 nm shows a negative deviation because the resonance shifts out of the pump. For FW = 1756 nm, the deviation remains minimal, around -0.02, due to the resonance moving out of the pump wavelength, although it is barely visible on the log-log scale. In addition, an oscillatory deviation is observed in the experimental data for incident powers up to 120 mW, a behavior not captured by the simulation. This phenomenon is likely associated with two-photon absorption and subsequent radiative processes under high-power excitation.

After describing these nonlinear generation effects under the TH regime, we further analyzed the dynamic influence of the fundamental mode on the THG response. In Fig. 4(F), the size of each circle represents the scattering cross section of different multipolar contributions: the red circle denotes the magnetic dipole, while the blue circle indicates the electric dipole. To isolate the impact of the mode from the power contribution, we normalized the scattering cross section to a common scale. Notably, the magnetic dipole decreases gradually with increasing incident power, whereas the electric dipole exhibits the opposite trend. This behavior provides a qualitative explanation for the variation of efficiency enhancement in THG. This analysis indicates that at low incident power, the magnetic dipole offers greater field enhancement than the electric dipole. Consequently, the THG at the magnetic dipole resonance wavelength surpasses that of the electric dipole, provided there are no other mode contributions at each THG wavelength. However, as the LC rotates, the magnetic dipole's field enhancement diminishes, while the electric dipole's field becomes stronger.

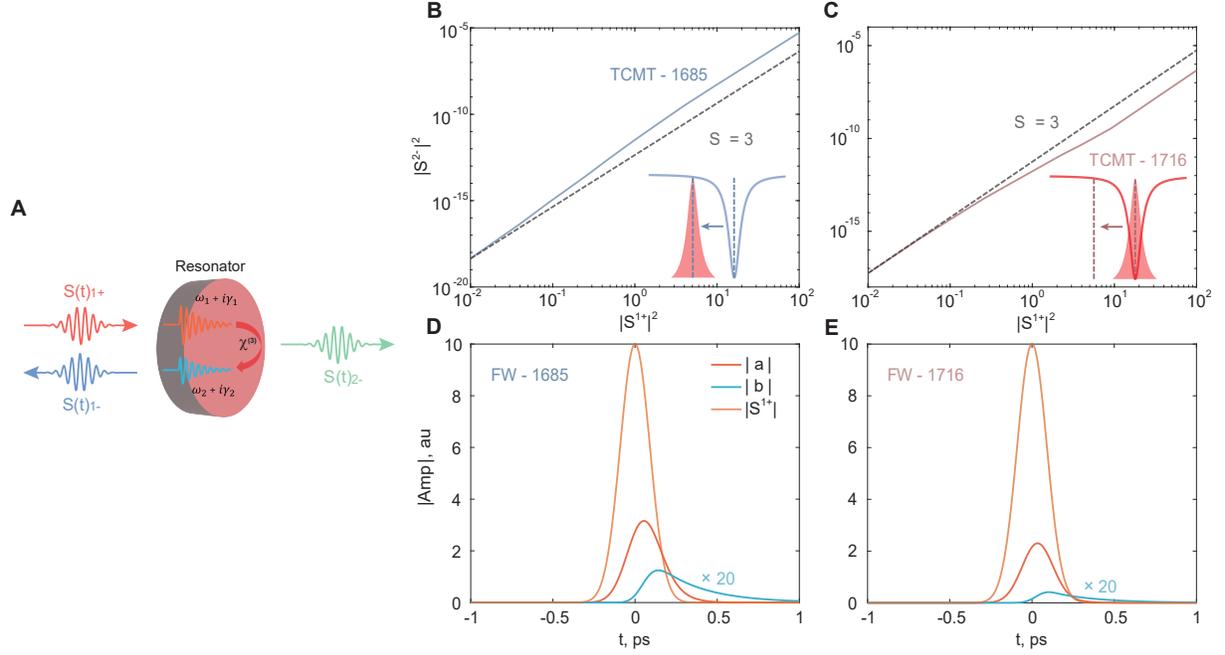

**Figure 5**: **Theoretical mechanism demonstration of bending curvature behavior in THG by temporal coupled mode theory (TCMT).** (**A**) Schematic of resonator nonlinear response. (**B**) illustrates the positive curvature bending due to the magnetic dipole resonance moving into the FW, and (**C**) shows negative curvature bending because the resonance is moving out of the FW. (**D**) and (**E**) are the amplitudes of mode $a$ and $b$ at incident pulse amplitude $|S^{1+}(t)| = 10$, where mode $b$ is magnified 20 times for observation.

## Non-cubic curvature mechanism analysis

In this section, we further analyze the observed non-cubic THG dependence by developing a nonlinear temporal coupled mode theory (NTCMT). This approach allows us to confirm the fundamental mechanism underlying the non-cubic dependence and also to provide insight into the time domain behavior in the THG process (*61, 62*). The NTCMT captures the dynamic time evolution of the resonant modes under external stimulation conditions.

To implement this model in our system, we consider three interacting modes: a magnetic dipole mode and an electric dipole mode at the fundamental frequency, and a nonlinear higher-order mode at $3\omega_i = 561.67$ or 572 nm. Because the electric dipole mode is orthogonal to the magnetic dipole mode for spherical structures, it exhibits weak and constant coupling in the disk-shaped nanostructure at the fundamental frequency. The model can therefore be reduced to two coupled equations. Terms corresponding to self- and cross-modulation are neglected, as their contribution is negligible given the weak nonlinearity of the metasurface compared to bulk nonlinear crystals.

$$\begin{aligned}\frac{da}{dt} &= -(i\omega_1 - \gamma_1)a + \sqrt{2\gamma_1}\, S^{1+} - i\left(\frac{\kappa}{3}\right)(a^*)^2 b, \\ \frac{db}{dt} &= -(i\omega_2 - \gamma_2)b - i\kappa a^3,\end{aligned} \qquad (4)$$

where $a$ and $b$ denote the amplitudes of the fundamental and third harmonic modes, respectively. A weak coupling between the two modes is assumed, with the system driven by a narrowband Gaussian pulse defined as $S^{1+}(t) = G(t)\exp[-i\omega_i t]$, where $G(t) = E_i \exp\left[-\frac{t^2}{2\Delta t^2}\right]$. Here, $E_i$ is proportional to the incident power, and $\omega_i$ is the incident frequency. In our model, the central frequency $\omega_1$ and the damping rate $\gamma_1$ of the fundamental mode depend on the rotation angle of the LC, $\theta$, while $\omega_2$ and $\gamma_2$ for the third harmonic mode are treated as constants for simplicity.

By extracting the absolute value of the output $|S^{2-}|$ from the calculation, we obtain the relationship between the input and the output power. The results are shown in Fig. 5. Fig. 5(A) illustrates the schematic of the nonlinear response of the resonator, where the fundamental and TH modes are coupled through the third-order susceptibility $\chi^{(3)}$, allowing bidirectional (forward and backward) propagation.

In this analysis, we consider only the forward coupling channel, consistent with the experimental measurements. Fig. 5(B) illustrates the resonance evolution as it enters the pump wavelength. The central wavelength of the fundamental mode gradually shifts and eventually aligns with the pump wavelength. As the resonance approaches the pump wavelength, the TH output exhibits a rising trend with increasing input power, followed by a constant slop once the resonance wavelength aligns with the pump wavelength. This behavior, consistent with the experimental observation in Fig. 4(A), provides a physical explanation for the up-bending curvature observed in THG. Conversely, if the resonance initially aligns with the pump wavelength and then gradually shifts away, as shown in the inset of Fig. 5(C), the system exhibits a down-bending curvature in the THG response. This trend also coincides with the experimental results in Fig. 4(B).

Moreover, we further analyze the time-domain evolution of the FW amplitude $a$ and the TH mode amplitude $b$ under the input amplitude $|S^{1+}(t)| = 10$. In the weak nonlinear coupling regime, the TH mode exhibits a much weaker yet clearly resolved temporal response (×20 for observation)

that temporally overlaps with the FW pulse. This behavior confirms that the TH field is generated parametrically from the instantaneous FW amplitude through nonlinear coupling, rather than being independently excited. The temporal localization of the TH signal near the peak of the FW pulse reflects the cubic dependence of the nonlinear source term on the FW amplitude, resulting in significant TH generation only when the cavity FW field is sufficiently strong.

Although the FW temporal profiles in Fig. 5(D) and (E) are similar, a pronounced difference is observed in the TH amplitude. This contrast highlights that the efficiency of harmonic generation is governed not only by the FW field strength but also by the spectral matching conditions of the harmonic mode. Together, these results reveal the fundamental physics underlying the nonlinear behavior of THG under external stimulus in our system. Furthermore, this simplified model provides both predictive numerical solutions and a clear physical interpretation of the nonlinear dynamics in complex photonic systems.

## Diffraction patterns modulation

In this section, we numerically demonstrate a direct application of the dynamic tuning of THG. Beyond modulating the resonance lineshape and conversion efficiency, our approach also allows for real-time control of the nonlinear diffraction pattern. This behavior arises because, in the nonlinear regime, the diffraction pattern is predominantly governed by the spatial profile of the nonlinear mode. By analyzing the far-field response derived from the TH near-field distribution, we can accurately obtain the diffraction pattern formed at the back focal plane. Simultaneously, the LC layer introduces a refractive index change in the nonlinear regime as well. As the LC rotates between different states, it perturbs the optical mode at the TH wavelength, causing a resonance spectral shift. Consequently, the diffraction pattern evolves continuously with the LC reorientation, as illustrated in Fig. 6.

In Fig. 6, we present results for the FF = 1685 nm and FF = 1716 nm, investigated in the previous section. Fig. 6(A) and (B) show the diffraction patterns for input power 10 mW and 100 mW at FF = 1685 nm. A distinct change is observed at the first diffraction orders ($m, n = \pm 1$), indicating the dynamic tuning effect. To guide the interpretation of the diffraction orders, we overlay a white circle in Fig. 6(A), (B), (D), and (F), which represents the critical angle for total internal reflection.

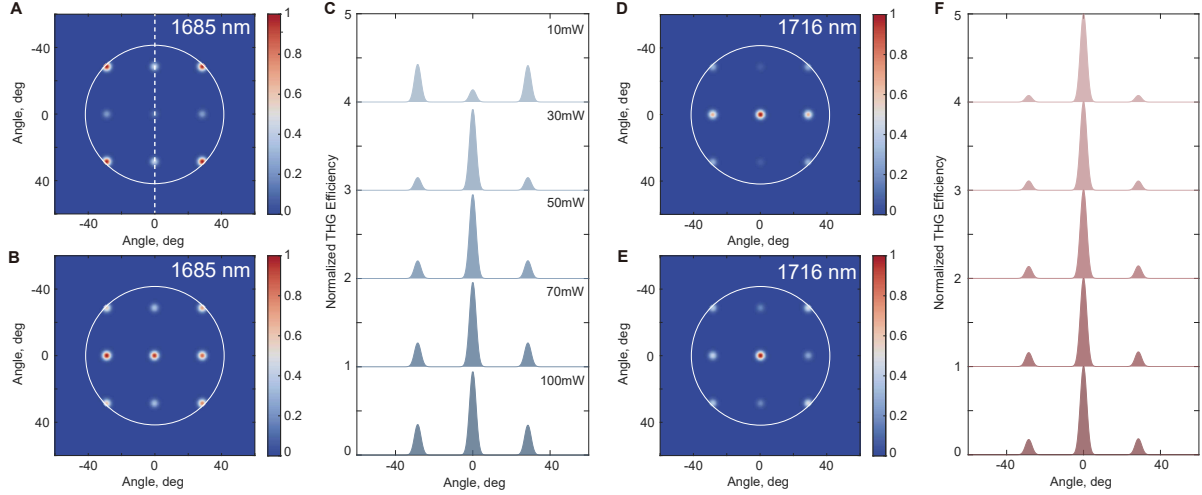

**Figure 6**: **Power-dependent modulated TH back focal plane intensity patterns** (**A**) and (**B**) are the simulated TH diffraction patterns at FW of 1685 nm with 10 mW and 100 mW pump power, respectively. (**C**) is the 1D intensity plot of a vertical dash line at (**A**) with different powers. A clearly pattern modulation from 10 mW to 30 mW due to a high-order mode shift away from $3\omega$ ($\omega = c/\lambda$, where $\lambda = 1685$ nm). (**D**) and (**E**) are the TH diffraction patterns for 10 mW and 100 mW pump power at FW of 1716 nm, and (**F**) is the 1D intensity plot from (**D**) to (**E**). Without being influenced by high-order modes, the TH intensity of 1716 nm demonstrates a limited modulation at order $m, n = \pm 1$.

Diffraction orders appearing outside this circle are confined within the substrate and do not radiate into the far field.

To quantify the modulation of each diffraction order, we extract a vertical cross-section along the dashed white line in the two-dimensional pattern and plot the normalized intensity distribution in Fig. 6(C). A pronounced enhancement of the zero-order diffraction is observed as the pump power increases from 10 to 30 mW. We note that this range refers to the simulated evolution of the normalized TH diffraction pattern, and this behavior is attributed to the mode shifting at TH wavelength, as confirmed by the corresponding TH field profiles shown in S7. The variation in the field distribution arises from the reorientation of the LC, which effectively alters the mode profile and, consequently, the far-field diffraction pattern. These results reveal that, at this representative TH wavelength $1685/3 = 561.67$ nm, the zero-order diffraction can be selectively suppressed or enhanced by tuning the incident pulse power. The TH pattern stabilizes above 30 mW, and further increases in power no longer affect the normalized intensity of the zero-order. Instead, it leads to a gradual enhancement of the first-order diffraction ($m, n = \pm 1$). This contrast suggests that the modulation of the diffraction pattern in the nonlinear regime arises primarily from nonlinearly generated resonance tuning at the TH wavelength. By varying the pump power to obtain the THG, we simultaneously drive LC reorientation, thereby enabling dynamic control over the far-field diffraction response.

Similarly, Fig. 6(D), (E), and (F) display the diffraction pattern variation at FF = 1716 nm. In this case, the rotation of the LC redistributes the relative weights of the dominant multipolar contributions at the TH wavelength, which in turn modifies the far-field diffraction patterns through the coherent interference of multiple electric and magnetic multipoles (see Supplementary S8 for details). As can be seen, the diffraction pattern remains unchanged across the range of pump powers, suggesting a consistent nonlinearly generated multipoles content with different pump powers. In this case, as shown in Fig. 6(F), the rotation of the liquid crystal causes only a slight variation in the $\pm 1$ diffraction intensities without altering the overall diffraction pattern. In contrast to Fig. 6(C), the modulation observed at 1685 nm arises mainly from the excitation of a resonant mode at the TH wavelength. Therefore, by engineering the mode state at the TH wavelength and varying the pump power, the optical field can simultaneously drive the liquid crystal to reorient and induce mode conversion at the TH wavelength, thereby achieving dynamic control over the far-field diffraction response.

## DISCUSSION

Our work establishes and validates a new modulation route, an all-optical route, to manipulate both linear and nonlinear responses of LC-infiltrated silicon metasurfaces. Starting from a realistic metasurface platform, simulations and multipolar analysis identify two overlapping resonances, mode 1 (magnetic dipole) and mode 2 (electric dipole), whose fields sample the LC environment differently. A continuum model based on the optical Freedericksz transition is employed to describe the pump-induced rotation of the LC molecules. The resulting orientation profiles are converted into an effective anisotropic refractive index using Euler rotation, allowing for the prediction of the resonance dynamics driven by PIOT. From a broader research perspective, spatially structured optical fields, particularly those with engineered polarization distributions, can generate spatially varying optical torque densities, potentially enabling programmable LC reconfiguration patterns beyond uniform intensity-driven actuation.

Linear measurements confirm a pronounced blue shift of the magnetic dipole mode (36 nm) and a minor red shift of the electric dipole mode (3 nm). The polarization response of the system rules out thermal origins and indicates an experimental LC-rotation threshold on the order of tens of milliwatts (0.94 k$W/cm^2$ average). In the nonlinear regime, power-dependent THG at three fundamental wavelengths exhibits characteristic "bending" in log–log plots. Namely, we observe an enhancement above the cubic power dependence when the resonance shifts into the pump. Conversely, we observe suppression of the THG below the cubic power dependence when the resonance shifts out of the pump wavelength. The static case follows the cubic scaling power law as a reference. Such tunable power-dependent polynomial nonlinear transfer functions in THG provide a promising route to reconfigurable optical nonlinearities for neuromorphic computational photonics, where adaptable activation functions and nonlinear mappings are central resources. After that, a phenomenological fitting model with power-dependent effective $\chi^{(3)}$, combined with nonlinear TCMT, successfully reproduces the observed trends and attributes them to resonance detuning induced by LC reorientation. In addition, the model captures the temporal response of the fundamental and TH modes, providing clear insight into the dynamical interplay between nonlinear excitation and resonant mode coupling.

Furthermore, far-field simulations reveal power redistribution within the TH diffraction pattern with varying LC orientations, thereby enabling selective enhancement or suppression of specific diffraction orders. Collectively, these results verify the feasibility of optical-torque-driven control of nonlinear emission in LC-assisted metasurfaces and establish a clear physical link between energy input, molecular reorientation, and nonlinear output. The established framework provides quantitative design guidelines and practical relevance for tunable frequency conversion, nonlinear imaging, and multifunctional meta-devices, paving the way for applications of optically driven reconfigurable nonlinear photonic systems.

# Acknowledgments

We thank S. Klimmer for his discussion on nonlinear fitting and D. Smirnova for her discussion on the nonlinear TCMT.

**Funding:** We acknowledge support from the Australian Research Council, Centres of Excellence program (CE200100010). This work was funded by the German Research Foundation (DFG) Meta-Active IRTG 2675 (437527638 and 490736786) and NOA CRC 1375 (398816777); the UK Research and Innovation Future Leaders Fellowship (MR/Z000270/1); and the European Research Council Consolidator Grants (UPIRI 101170298). Views and opinions expressed are, however, those of the authors only and do not necessarily reflect those of the European Union or the European Research Council. Neither the European Union nor the granting authority can be held responsible for them. The authors acknowledge the use of NTU High-Performance Computing cluster Avicenna.

**Author contributions:** Z.Y, L.X, M.R., and D.N.N conceived the idea and designed the research. Z.Y performed the theoretical studies. A.B. and Z.Y. developed, fabricated, and filled the LC cell. Z.Y, A.B. and C.B. designed the LC cell and alignment geometry. K.T. and I.S. supervised and fabricated the metasurface. Z.Y, M.Y., and S.L. provided the images of the metasurface. Z.Y. and G.S. performed the linear and nonlinear characterization experiments. Z.Y., L.W., M.L., L.X., T.P., and D.N.N analyzed and discussed the results. Z.Y. wrote the paper with input from all authors.

**Competing interests:** There are no competing interests to declare.

**Data and materials availability:** All data needed to evaluate the conclusions in the paper are present in the paper and/or the Supplementary Materials.